\documentclass[prl,twocolumn,groupedaddress]{revtex4}
\usepackage{graphics,epsfig}

\begin{document}


\title{How does Europe {\em Make Its Mind Up?} \\ 
Connections, cliques, and compatibility between countries  \\
in the Eurovision Song Contest 
}
\author{Daniel Fenn$^a$, Omer Suleman$^a$, Janet Efstathiou$^b$ and Neil F. Johnson$^{a,1}$} 
\address{$^a$ Physics Department, Oxford University, Oxford OX1 3PU, U.K.}
\address{$^b$ Department of Engineering Science, Oxford University, Parks Road, Oxford OX1 3PJ, U.K.}

\date{\today}

\begin{abstract} We investigate the complex relationships between 
countries in the Eurovision Song Contest, by recasting past voting data in terms of a dynamical network. Despite the British tendency to feel distant from Europe, our analysis shows that the U.K. is remarkably compatible, or `in tune', with other European countries. Equally surprising is our finding that some other core countries, most notably France, are significantly `out of tune' with the rest of Europe. In addition, our analysis enables us to confirm a widely-held belief that there are unofficial cliques of countries -- however these cliques are not always the expected ones, nor can their existence be explained solely on the grounds of geographical proximity. The  complexity in this system emerges via the group `self-assessment' process, and in the absence of any central controller. One might therefore speculate that such complexity is representative of many real-world situations in which groups of `agents' establish their own inter-relationships and hence ultimately decide their own fate. Possible examples include groups of individuals, societies, political groups or even governments. 

\vskip0.2in
 
\end{abstract}

\maketitle

\footnotetext[1]{Address for correspondence:\\ 
Prof. Neil F. Johnson \ \ \ \ \  n.johnson@physics.ox.ac.uk \\ 
Physics Department, Clarendon Laboratory, Parks Road, Oxford OX1 3PU, U.K.\\ 
Tel: +44 (0)1865 272287  \ \ \  Fax: +44 (0) 8701 344065.}

\begin{figure}[ht] 
\includegraphics[width=.46\textwidth]{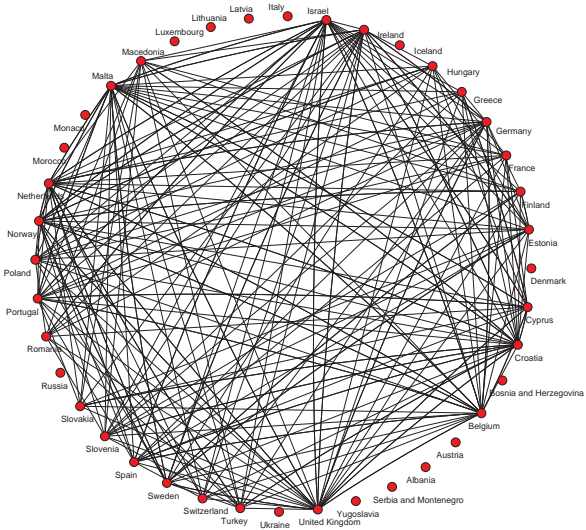}
\caption{The undirected, unweighted points network for the Eurovision Song Contest in 1998. Each country awards points to ten other countries, making the resulting network rather densely connected. The links either exist (i.e. points were awarded) or not (i.e. no points were awarded). There is no weighting to reflect the specific number of points.  The countries with no links at all are the ones not present in the competition in 1998.
}
    \label{fig:figure1} 
\end{figure} 

\noindent{\bf{1. Introduction}}

Europe seems to exhibit all the characteristics of what natural scientists call a {\em Complex System} \cite{nino}. It consists of a network of many objects (i.e. countries) whose complicated interactions depend both on geography and past history. In short, these interactions are non-local in both space and time. It is also a self-organized system, operating through collective decision-making. While  governments and industries may be keen to strengthen ties with their European counterparts, popular opinion seems rife with prejudices and suspicions on the subject. Such wariness is not too surprising: Europe has, after all, succeeded in giving birth to two World Wars within less than one hundred years. Indeed, it is remarkable that a single European constitution could soon emerge only sixty years after the end of World War II. Such an emergent phenomenon is itself a defining characteristic of a Complex System \cite{nino}.

Some countries are traditionally seen as `pro-European' while others are labelled `Euro-skeptic'. In the case of the UK, for example, frequent media headlines such as `Britain shuns EU info centres'  \cite{guardian} arguably fuel the country's Euro-skeptic image. Referenda on the EU Constitution are now taking place in individual countries -- for example, France will vote on 29 May 2005. However it has been reported that most people are unaware of the precise content of the proposed constitution \cite{bbc}. Given this lack of factual knowledge, it seems likely that any widespread Euro-skepticism among the population of a given country A is merely reflecting A's implicit distrust of how countries B, C, D {\em might} act within a Europe-wide system. Going further, it seems possible that any such distrust is   
driven by a fear that B, C, D are fundamentally  {\em incompatible} with A. Hence a citizen of A perceiving such an incompatibility with countries B, C, D, might therefore consider it a bad idea for A to permanently join up with B, C, D. After all, people pay considerable attention to such incompatibility issues when deciding on potential partners in their private lives.

This discussion raises an interesting question: What phenomenon could we measure in order to examine how compatible the various European countries actually are? Just as in the physical sciences, our ideal phenomenon would be measurable in a transparent, quantifiable and repeatable way, which implies repeating exactly the same experiment in all countries under the same conditions and over many years. However it should not be so economics-driven that it ends up depending on commercial or political issues. The required phenomenon must capture some broad section of public sentiment and tastes, rather than those of elite sectors. It should therefore be something which is universally understandable and which automatically engages the interest of the majority of the population within each country. It should not depend directly on individuals' income, education, or language -- instead, it should capture something fundamental about the underlying character and mood within a given country at any given point in time. In short, we are looking for a phenomenon whose measurable values might allow us to deduce how `in tune' different countries have been in the past, are in the present, and might be in the future.

On 21 May 2005, the Eurovision Song Contest celebrates its 50th anniversary \cite{ref1}. This 2005 contest will probably attract the largest global television and radio audience of the year -- the combined television and radio audience figures in recent years have approached one billion. Although the specific rules of the contest have changed over the years, the basic format is the same: Each participating country performs a song, and this song is then awarded points by other countries. 
Irrespective of whether it contributes anything to the advancement of music per se,  the Eurovision Song Contest does provide a remarkable and unique example of an annual exchange of `goods' and opinions between countries. Going further, it is arguably the only international forum in which a given country can express its opinion about another, free of any economic or governmental bias. Indeed if we assume for the moment that a given song either sounds `nice' or not \cite{ref13}, then it should receive the same order-of-magnitude of vote from all countries. Hence any large differences in voting may be reflecting some deeper sociological differences between countries. Assuming that all countries have equal chances of producing intrinsically `nice' songs over the timescale of a decade \cite{ref13}, then any systematic bias which arises in the voting patterns of country A toward countries B, C and/or D may be telling us something about how compatible A is with B,C and/or D. In this sense, the voting in the Eurovision might be regarded as the sociological equivalent of The Economist's {\em Big Mac Index} which compares the measured value (i.e. cost) of a particular product within different countries \cite{ref2}. It has even been suggested that the concept of the Eurovision Song Contest as a whole should be used as a role-model for determining the overall composition of the European Union \cite{ref3}.

In this paper, we use the framework of complex dynamical networks \cite{ref4,ref5,ref6,ref7,ref8,ref9,refus1} in order to analyze voting behaviour in the Eurovision Song Contest over space (i.e. between countries) and time (i.e. between years). Although some previous studies of the Eurovision Song Contest do exist \cite{ref10,ref11,refnew,ref12}, our study is unique for the following reasons: (a) We analyze the voting data from the point of view of a complex evolving network. This enables us to uncover non-trivial, non-linear patterns from a large amount of `noisy' data. (b) We look across multiple timescales, focusing on the patterns which emerge between years. (c) We look across all countries regardless of whether they won or not. (d) We consider data over the recent period 1992-2003 inclusive, during which the number of countries participating is fairly constant. (e) Our analysis focuses on the points given and received by all countries, rather than the final outcome of the contest. 

Some of our conclusions serve to confirm several commonly-held beliefs about particular  cliques of countries. However we also uncover some very surprising and unexpected results. In contrast to some commonly-held beliefs  -- in particular within the U.K. itself -- the U.K. has been consistently `in tune' with the rest of Europe since the early 1990Õs. Just as surprising is the fact that France, for example, has been rather `out of tune' with the rest of Europe over the same period \cite{ref16,2003}. [N.B. We will take the term ÔEuropeÕ to include all the countries participating in the Eurovision Song  Contest. These include, for example, Israel which has actually won the contest three times. Other countries in the Middle East are apparently also keen to join the contest sometime soon. Repeating our study in a decade's time could therefore provide some even greater surprises.] 
Our analysis is built around the framework of complex networks \cite{ref4,ref5,ref6,ref7,ref8,ref9} and focuses only on the Eurovision Song Contest. However we note that the analysis tools that we have used, and those that we have introduced in the course of this work, have more general applicability. This is because the Eurovision Song Contest, in its most elementary form, is just an example of a set of entities repeatedly exchanging some goods. In another context, these goods could equally well be ideas, opinions, money, supplies, food or nutrients. In this sense, our analysis should also be applicable to time-evolving network systems in sociology, biology, economics, business and even financial markets \cite{refus2}. For example, the World Trade web in Ref. \cite{ref8} takes a similar form to the networks which we create for the Eurovision Song Contest, but with countries exchanging goods rather than points.

\vskip0.4in
\noindent{\bf{2. The Eurovision Song Contest and the data}}

The idea of creating a Eurovision Song Contest was inspired by the popular Italian San Remo Festival. The contest was originally titled ÔThe Eurovision Grand PrixÕ, and had only seven participants when it started in 1956. Since then, the number of participants has risen steadily and currently exceeds twenty \cite{ref1}. The rules concerning the procedure for voting and the language in which the song must be sung, have also evolved over time. The basic structure of the current scoring system has been in place since 1975, whereby each voting country awards ten other countries. Specifically, each voting country A allocates the set of points $\{1, 2, 3, 4, 5, 6, 7, 8,10,12\}$ to the ten countries $\{{\rm B, C, D, E, F, G, H, I, J, K}\}$ that it considers as `best', where $\{{\rm B, C, D, E, F, G, H, I, J, K}\}$ are a subset of the entire set $\bf S$ of competing countries. Countries are not allowed to vote for themselves. Although its name refers to `Europe', the Eurovision Song Contest regularly includes countries such as Israel, while other traditionally non-European countries are also petitioning to enter. Regardless of one's opinion about the Eurovision Song Contest itself, the quantity of countries competing and the number of television viewers across the globe suggest that it is one of the world's few truly international events.

Given the way in which the points are awarded, and the fact that they are recorded in a database on the organizersÕ official website \cite{ref1}, one can easily construct  a network for each year of the contest by representing the countries as nodes and the points exchanged as edges, acting as connections between the nodes. This network can either be directed (e.g. with directed edges drawn from the country giving points to the country receiving points), undirected (e.g. with undirected edges drawn between countries if points are exchanged in either direction), weighted (e.g. with edges drawn such that the weight of an edge is equal to the number of points exchanged), or unweighted (e.g. with edges drawn such that the actual number of points is ignored). Since each country assigns points to ten other countries, the minimum degree that a country can have is ten and the maximum degree is equal to the total number of other countries in the competition in a particular year. An example network is shown in Fig. 1. We emphasize that the type of network analysis which we present in this paper is not limited to the Eurovision Song Contest -- the same ideas can be applied to any type of network in which the links between nodes represent exchanges. 
An interesting aspect of a Eurovision Song Contest points-based network, as compared to many other networks which are typically studied in the literature, concerns its temporal evolution. To understand this more clearly, consider a standard collaboration network between scientists \cite{ref7}. The scientists, represented by the nodes, are considered connected if they have authored a paper together. At an initial point in time, the network will exist with a number of links between scientists. As time passes, more links will appear as new collaborations are made and as new nodes -- corresponding to scientists publishing for the first time -- appear. The original connections will, however, persist. Particular nodes will stop gaining new edges as scientists cease publishing. By contrast to this continuous evolution, the Eurovision network evolves discretely. The network for each point in time is in essence an isolated entity with limited correlation between years. In any given year, links will exist between countries if they vote for each other. The next year it is not necessary for any of these links to remain since countries can change who they assign points to. A further complication results from the fact that the nodes change since different years may contain different competitors. If a country connected by an edge does not compete in consecutive years, then it is impossible for a link to persist.
The notable network properties should come to light through a comparison of these annual points-based networks. Between most pairs of consecutive years, roughly eighty percent of the countries will be common and so it is interesting to investigate which, if any, links persist. This leads to one of the main avenues of investigation in the present paper.

The fact that there have been a number of changes to the format of the Eurovision Song Contest over the years, means that great care must be taken with the analysis in order to deduce meaningful results regarding possible space-time patterns in voting behaviour. The key changes which led us to the particular analysis which we employ in this paper, are as follows: (i) In 1975, the current scoring system was introduced in which countries award 1-8, 10 or 12 points. This implies that if the weights of links are to be analyzed consistently, one should only consider networks produced after 1975. (ii) In 2004, a new format was devised with the competition being split into a semi-final and a final. From 2004 onwards, all countries involved in the semi-final can vote in the final, but only those that reach the final can receive votes. The introduction of a semi-final in 2004 allows the inward degree of a node to be significantly greater than in previous years, thereby making the results from 2004 onwards structurally different and potentially far more complicated. (iii) The number of countries entering the competition between 1975 and 2003 fluctuated between 18 and 26. However, in 1992 this number stabilized at around 23, with the number of countries competing in the period 1992-2003 varying only slightly between 23 and 26. Given these three considerations, all the subsequent analysis is therefore performed over networks produced in the stable period 1992-2003 inclusive \cite{2003}. It turns out that thirty-six different countries compete in this period 1992-2003, but only twenty-six appear for seven years or more. The investigation will therefore focus on these twenty-six regular performers.

\vskip0.3in
\noindent{\bf{3.	Analysis of the voting network and its comparison to a `random contest'}}

\vskip0.1in

\noindent{\em{3.1 Background}}

The network analysis will be performed on two scales: global and local. An initial investigation of the global properties of the network will be undertaken, before shifting to a local-scale consideration of the properties of individual nodes and edges. As will be shown, these two types of analysis provide important but complementary insights. When considering individual nodes, it is a relatively straightforward matter to draw a picture of the network with actual points and lines showing the interactions of the particular country, and to answer specific questions about network structure by examining this picture (see Fig. 2). However when considering the network of all nodes for a particular year (as in Fig. 1) this approach is practically useless. It is therefore necessary to use statistical properties, such as degree distributions, that characterize the structure and behaviour of the network.

In order to gain some perspective on the results obtained from the empirical Eurovision data, it is necessary to have something to compare them to. An insightful comparison can be made by comparing it to the results one would obtain if the Eurovision Song Contest were a `random contest'. In order to simulate such a `random contest', we will assume that each song possesses the same fundamental value (i.e. all songs are equally `nice' \cite{ref13}). Hence all songs will be equally attractive to country A. The `random contest' model also needs to assume that a given country A possesses no biases in terms of wanting to favour or penalize countries B,C,D, etc. Hence the country of origin of a given song does not affect the probability that country A will vote for it.  Given these two assumptions, country A will simply assign its ten packets of points (i.e. 1-8,10,12) randomly  among the remaining countries. Any links between countries as a result of the points cast, will therefore exist independently of each other. If there are $N+1$ countries in a particular year including country A, then the probability that country A votes for another country B in this `random contest' is given by $p_{A\rightarrow B}=10/N$. The $10$ in the numerator results from the fact that each country can assign points to ten other countries. We will use this `random contest' construction in order to generate surrogate data which can then be compared to the actual Eurovision data. This will then help us uncover any significant (i.e. non-random) patterns in the actual Eurovision data.

\begin{figure}[ht] 
\includegraphics[width=.48\textwidth]{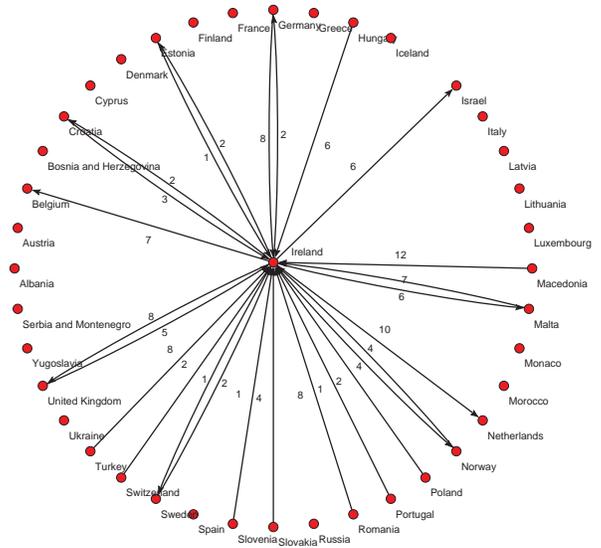}
\caption{An example showing the network as seen from the point of view of a single node. This case shows the network for Ireland in 1998. The numbers shown on each line indicate the points awarded by, and to, Ireland. The arrowheads give the direction of the link e.g. Ireland gave UK 5 points while it received 8 from UK.
}
    \label{fig:figure2} 
\end{figure} 

\vskip0.2in

\noindent{\em{3.2  Clustering Coefficients}}

In most social networks \cite{ref7} two nodes that are linked to a third node have a higher probability of having a link between them: acquaintances of a given person are more likely to know each other. It is possible that the same effect could be observed in the Eurovision network. If, as is often hypothesized, there are a number of voting `cliques' within the contest, then it is reasonable to assume that the observed clustering coefficients \cite{ref3} will be greater than those arising from the `random contest' surrogate data. In short, if two countries both vote for a third country, then there would be a higher probability that they also vote for each other. The clustering coefficient $C$ of a particular vertex is defined as the probability that two neighbours of a given vertex are also neighbours of one another. It then follows that  $C$ can be calculated by evaluating the number of edges between the neighbours of a given vertex $\nu$, and dividing this quantity by the combinatorial quantity $d(\nu)!/2! (d(\nu)-2)$!  where $d(\nu)$ is the degree of vertex $\nu$. The denominator is just the maximum possible number of edges between the neighbours of $\nu$. This implies that $0\leq C \leq 1$. Averaging $C$ over all vertices of a network yields the clustering coefficient of the network. 
Using this quantity, we have calculated the clustering coefficients for each annual network as in Fig. 1, between 1992 and 2003. Throughout this paper, we have used the software program Pajek to help with the visual representation and analysis of our results \cite{ref14}.

For the Ôrandom contestÕ (i.e. random network) the expected value of the clustering coefficient is equal to the probability that two randomly selected nodes are connected. This results from the fact that the links exist independently and so the probability that two countries vote for each other is not affected by the fact that they both also vote for a common neighbor. To find the clustering coefficient for this randomized version of the Eurovision data (i.e. a random network) it is therefore necessary to calculate the probability that a link exists between two countries. Assuming that there are $N+1$ participating countries in a particular year, then the probability that country A votes for country B, or vice versa, is given by:
\begin{equation}
p_{A\rightarrow B} = p_{B\rightarrow A} = \frac{10}{N}\ \ .
\end{equation}
The probability that there is a link between two countries will then be:
\begin{equation}
p_{\rm link} = p_{A\rightarrow B} + p_{B\rightarrow A} - p_{(A\rightarrow B) {\rm AND} 
(B\rightarrow A)}
\end{equation}
which, given the independence of the links, can be regrouped to give
\begin{equation}
p_{\rm link} = 2 p_{A\rightarrow B} - [p_{A\rightarrow B}]^2
\end{equation}
and hence the clustering coefficient for the `random contest' is given by
\begin{equation}
C_{\rm random} \equiv p_{\rm link} = \frac{ 20 (N-5)}{N^2}\ \ .
\end{equation}

\begin{table}[ht] 
\includegraphics[width=.51\textwidth]{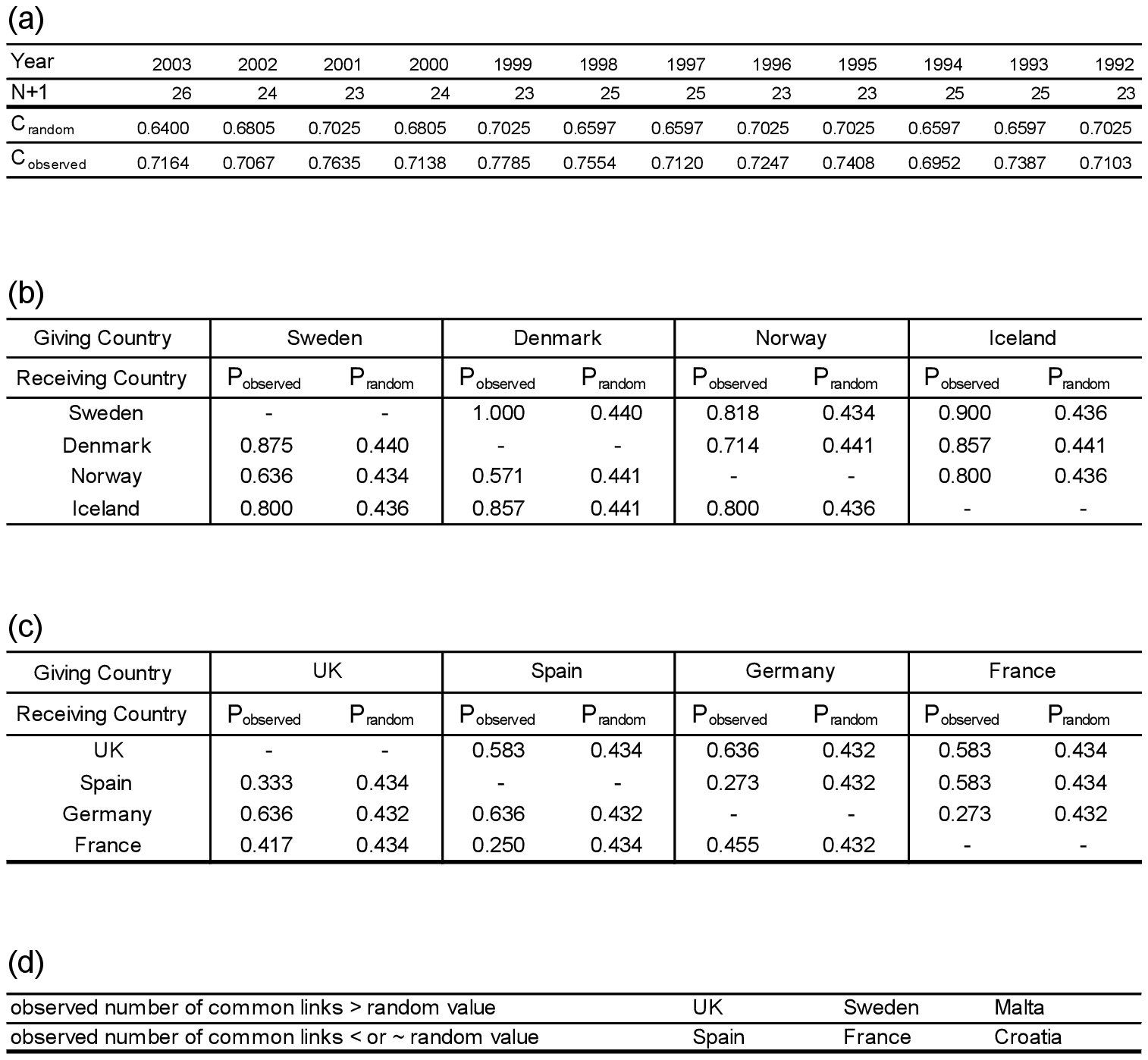}
\caption{(a) The observed clustering coefficient $C_{\rm observed}$ and `random contest' clustering coefficient $C_{\rm random}$, for 1992-2003. There are $N+1$ participating countries.
(b)  The observed probability $p_{\rm observed}$ and the Ôrandom contestÕ probability $p_{\rm random}$, that the Nordic countries assign points to each other. The observed probability $p_{\rm observed}$ is deduced from the actual data. The countries giving points are shown along the horizontal, while the countries receiving points are shown down the vertical. (c) Same as in (b) but for the Big Four group. (d) An approximate categorization, organized in terms of the observed number of common links, of the six countries that compete in the competition for eleven or twelve consecutive years between 1992 and 2003. The top line shows the countries for which the observed number of common links is typically  greater than the random value (i.e. UK, Sweden and Malta). The bottom line shows the countries for which the observed number of common links is typically less than or equal to the random value (i.e. France, Spain and Croatia).
}
\end{table} 

Table 1(a) and Fig. 3 show the observed and random values of the clustering coefficients. As can be seen, the observed clustering coefficients are always greater than the random graph values. The Eurovision network contains, on average, around twenty nodes and each node must be connected to at least ten other nodes. This small network size, combined with the high probability of connection, means that one would always expect a high clustering coefficient for a network of this nature. However, the results do  provide an indication that the system is not random and that there are indeed some voting `cliques'.

\begin{figure}[ht] 
\includegraphics[width=.48\textwidth]{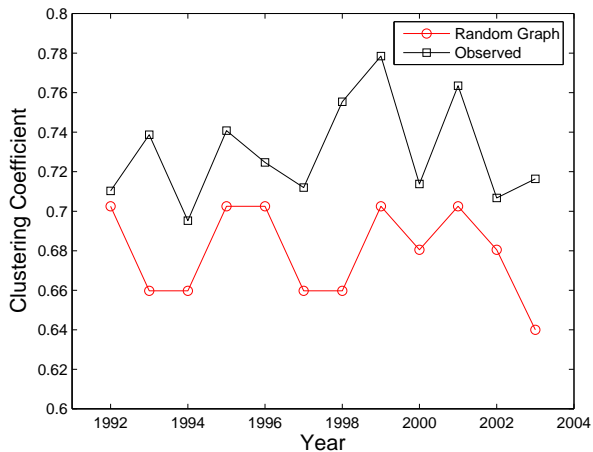}
\caption{Comparison of the observed clustering coefficient for the actual Eurovision data for 1992-2003, and the corresponding random graph result for the `random contest' (see text for description).
}
    \label{fig:figure3} 
\end{figure} 

\vskip0.2in

\noindent{\em{3.3  Degree Distribution}}

The degree distribution of the nodes provides further evidence that the Eurovision network is not random. We recall that the degree of a vertex is the number of edges connected to that vertex. For the Eurovision network, there are various possible definitions of the vertex degree depending on whether in, out or in-and-out (i.e. reciprocal) connections are considered. In this section, the degree of a vertex will be taken to mean the number of edges connected to that vertex. No distinction will be made between inward, outward or reciprocal connections.

	For any particular year, there are only around twenty nodes and so it is difficult to draw any reliable conclusions about the degree distribution. In order to increase the size of the data set, the degree of nodes for all of the years between 1992 and 2003 will be considered. Since the number of nodes between years differs, it is necessary to normalize the degree values so that the data between years is equivalent. This is done by dividing the degree of each node by the total number of other countries appearing in that year. Instead of viewing a degree distribution, one will then be looking at the distribution of degree as a fraction of the maximum possible degree. This quantity is universal across years and will produce a distribution that is homologous with the degree distribution over all years. One can then define $P(k/N)$ to be the overall fraction of vertices in the networks that have a normalized degree $k/N$. Equivalently, $P(k/N)$ is the probability that a vertex chosen uniformly and at random, has a normalized degree equal to $k/N$. A plot of $P(k/N)$ can be produced by forming a histogram of the normalized degree of vertices. This histogram is the normalized degree distribution for the combined networks and is equivalent to the actual degree distribution. For the random network case, each edge is present or absent with equal probability. The resulting distribution describing the probability of having a degree $k$, has the usual binomial form \cite{ref1} which depends on  the average probability that a connection exists in a particular year, and the average number of countries appearing.

	Figure 4 shows the normalized degree distribution for the observed Eurovision Song Contest data. Far from being a binomial distribution, the actual distribution is right-skewed. The number of countries with a degree close to the minimum value of ten is significantly greater than the values expected for a random graph. An explanation for this behaviour lies in the nature of the vertices. If a country possesses a large number of reciprocal links then, if the nature of the links is not considered, its degree will be lower than if it had only in and out links. The results suggest that a large number of reciprocal relationships could exist between countries. This in turn suggests a degree of cliquishness, since particular pairs of countries will therefore be exchanging votes.

\begin{figure}[ht] 
\includegraphics[width=.48\textwidth]{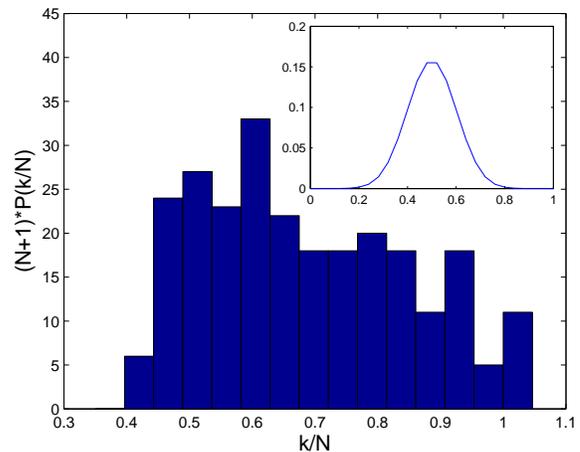}
\caption{Histogram showing the observed normalized degree distribution for the Eurovision Song Contest in the period 1992-2003, and (inset) the degree distribution which would arise for the `random contest' model. The x-axis is rescaled to the range $0\rightarrow 1$ in order to highlight the differences between the two curves.}
    \label{fig:figure4} 
\end{figure} 

\vskip0.2in

\noindent{\em{3.4 Cluster Analysis}}

The last two sub-sections have focused on global properties of the network. It is now instructive to perform a local-scale analysis by considering the properties of individual nodes and edges. In particular, we will use the notion of cluster analysis, whereby we will group together countries which behave in a similar way and hence can be regarded as being `close'. 

As a measure of each country's actions, we form a data series consisting of the average number of points assigned to each other entrant in the years in which they both compete. The closeness of each pair of countries can then be measured by comparing these data series using Pearson's correlation coefficient \cite{ref15}. The Pearson coefficient takes values ranging from $-1$ to $+1$. If two particular countries assign exactly the same number of points to each participating country, and thus possess identical data series, their Pearson coefficient will be one. The Pearson coefficients are then rescaled to produce a `distance' between $0$ and $2$ using the relationship that the rescaled distance  is equal to $\sqrt{[2(1-P_C)]}$ where $P_C$ is the Pearson coefficient. The most closely related countries have rescaled distances close to $0$, while the least correlated countries have distances close to $2$. This data is then used to plot a dendrogram which provides a visual aid for identifying clusters. 

Figure 5 shows the resulting dendrogram obtained by consecutively linking the most correlated countries. For example, Greece and Cyprus have the smallest rescaled separation and so they are combined first. The next smallest rescaled distance is between Denmark and Sweden and so they form the next cluster. Once two countries A and B have been combined into a cluster, they are considered to be at the same distance from another country C, which is equal to the shorter of the distances AC and BC. This construction is then generalized up for clusters with more than two countries. The distance between any two clusters is the shortest distance between any two countries in the two clusters. Progressively more countries and clusters are combined in this way, with some countries combining with existing clusters, until all the countries are united into a single cluster.

The dendrogram shows quite explicitly that the voting patterns of certain countries are highly correlated. Greece and Cyprus have a very small rescaled distance which demonstrates a very strong voting correlation. This particular finding thereby confirms a long-held belief among regular Eurovision viewers. There is a slightly less correlated cluster that involves the Nordic countries, Denmark, Sweden, Iceland, Norway, Finland, and somewhat surprisingly, Estonia. Many other clusters also arise: Bosnia and Turkey, Croatia and Malta, UK and Ireland (who also show a correlation with the Nordic clique), Belgium and the Netherlands and France and Portugal. The high correlations between the way in which countries assign points provides evidence in support of the theory that voting groups exist. Moreover many of the observed correlations may be dominated by only a few points in the data series which correspond to points that are assigned to countries within a particular voting clique.

Given the make-up of these groups, one may be tempted to conjecture that the correlation between countries' voting patterns is linked to their geographical closeness. If one considers that geographical closeness leads to cultural closeness, then this conclusion would seem reasonable since it is likely that neighbouring countries will have similar cultural tastes and sociological ties, and will therefore support each other's songs. This conclusion is, however, drawn into question by the existence of notable exceptions to the rule. Consider, for example, the exclusion of Spain from the group including France and Portugal; the presence of Estonia in the quasi-Nordic clique; the lack of correlation between Cyprus and Turkey. This suggests that the observed voting similarities in the dendogram, have arisen for rather more subtle reasons (e.g. a common, or uncommon, past history) as opposed to simply being the result of geographical proximity. 
 
\vskip0.2in

\noindent{\em{3.5  Network-based Analysis of Clusters}}

The statistical analysis in the previous section considered the average number of points assigned by countries. It is possible, however, that the results could have been distorted by a participant allocating an uncharacteristically large number of points to a country in a particular year. This leads us to pose the question as to what it actually means for two countries to be `close'. Is it most important to look at the similarities between how countries allocate all of their points, or should we only consider the number of points that particular countries assign to each other? The previous section focused on the former, but this also made it possible to conjecture on the latter. In the remainder of this paper, we will focus on this exchange of points within smaller groups of countries and in particular on the points exchanged by certain pairs of participants. Indeed, the subsequent analyses will neglect the weight of the links and will simply focus on the existence, or not, of directed links between countries. We will begin with the clusters determined in the previous section.

Using the individual node networks (see Fig. 2) it is simple to find the number of directed links that exist between two countries and then to convert this into a probability that a country will give points to another country in a particular year. The observed probability for the existence of a directed link $p_{\rm observed}$ is obtained by dividing the number of years in which country A gives points to country B, by the number of possible years in which points can be given. It is again instructive to compare these observed values with those derived for a random graph corresponding to the `random contest'.

\begin{figure}[ht] 
\includegraphics[width=.48\textwidth]{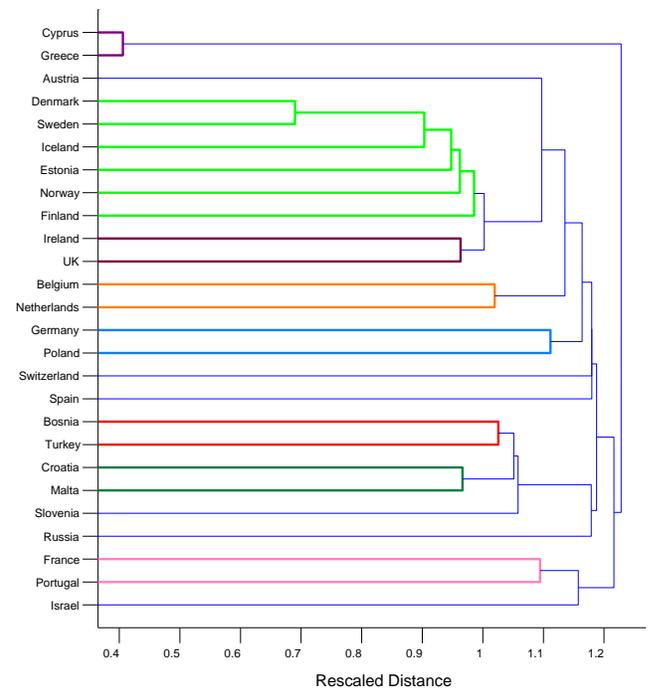}
\caption{Dendrogram showing the voting clusters within the Eurovision network. As the rescaled distance increases, the clusters become less correlated. Greece and Cyprus, for example, form the most correlated cluster.
}
    \label{fig:figure5} 
\end{figure} 

Consider two countries A and B. The number of countries competing varies between years and consequently so does the probability that country A assigns points to B in a particular year. In order to calculate a single probability that a country A will assign points to B, it is necessary to average over all the years in which both A and B compete. The resulting values can then be compared to the observed probabilities that a connection exists. We have performed this analysis for all possible clusters, but will only illustrate the results here for the four countries in the Nordic cluster, and also the group consisting of the `Big Four' countries Ð UK, Spain, Germany and France. This latter group is so named because they automatically enter into the final each year as a result of the television revenues which they generate.

Tables 1(b) and (c) show the observed and expected probabilities for the two clusters. Table 1(b) provides further supporting evidence that the Nordic group of countries form a voting clique. In all cases the observed probabilities that countries within the cluster assign points to other countries within the group, are higher than the corresponding probabilities for a Ôrandom contestÕ. Indeed, this discrepancy is rather marked in the majority of cases.
It is interesting to compare these results to the equivalent values for the Big Four group. The cluster analysis performed earlier did not indicate that these Big Four countries formed a cluster -- however they provide a useful comparison since they are present in nearly all of the years 1992-2003. Table 1(c) shows that the observed probabilities are often similar in magnitude to the random probabilities and that there are roughly equal numbers of cases when the observed probabilities are above or below the random value. This suggests that the Big Four group act toward each other as if it were a `random contest', and hence their internal voting network can be reasonably described as a random graph.	
These results are consistent with the earlier finding of no clique for the Big Four. 
In short, the voting patterns within the Big Four group of countries is much closer to random than for the Nordic clique. We have carried out this same analysis for all of the other clusters suggested by the cluster analysis. For each cluster, the probabilities that countries within the cluster exchange votes, are significantly higher than those predicted by the `random contest'.  These findings, while not shown explicitly here because of space constraints, confirm the internal consistency of our analysis and results so far.

\vskip0.2in

\noindent{\em{3.6  Common Link Analysis}}

Section 3.5 compared the probabilities that countries within particular groups vote for each other. This section generalizes this analysis by measuring the {\em persistence} of links between countries over time. In particular, we examine the number of consecutive years in which a link of any variety exists between two nodes. For simplicity, we only consider the presence or absence of a link between participants and do not distinguish between in, out or reciprocal links. This analysis allows us to identify possible `stable relationships' between countries. After all, if these were human relationships and we observed that two people regularly exchanged gifts or acts of kindness and affection, we would indeed suspect that these people had a stable relationship and hence would consider them `compatible'. 

For the purposes of discussion, consider first the case of two countries: A and B. If we find that a given country A either gives and/or receives points from another country B each year over a number of years, and that this exceeds that expected for a `random contest', then we label A and B as being compatible. Such compatibility implies that irrespective of any changes in the external world, A and B continue to identify with each other in consecutive years and hence interchange points. Likewise, if A and B were to change tastes radically, they would change in the same way such that they still continue to interchange points. Such compatibility will show up as a larger number of common links than in the `random contest'. If instead the reverse is true, and the number of common links over time is smaller than the expected `random contest' number, we would label A and B as `incompatible'.  Carrying out this common link analysis between country A and all other countries B,C,D etc., will give us a measure of how compatible country A is with the rest of Europe. Repeating this exercise for another country B, we can then deduce which of the two countries appears more compatible with the rest of Europe.  In a similar way, we can also deduce {\em incompatibility} between a given country and the rest of Europe, in the case that the number of common links is less than the `random contest' result.

	The observed numbers of common links over 2, 3, 4, 5, 6 and 7 consecutive year periods were found by analyzing the individual node networks (i.e. Fig. 2) for each of the countries which regularly participate. We also investigated whether the observed results can be reproduced, within standard errors, by the simple `random contest' model. When producing the corresponding random graph data, it is essential to consider the countries participating in a given year since non-participation significantly limits the number of consecutive years over which connections can persist. Of the twenty-six regularly participating countries, only five compete in each of the years 1992-2003. In order to produce results for a corresponding Ôrandom contestÕ, a program was written to randomly allocate points to countries: Each year was represented by a receiving-country/giving-country matrix with each non-participant assigned blank entries and each participant assigned a unique random number. The columns were then sorted by size. The rows in each column containing the ten highest numbers were assumed to be the countries that had received points, and these values were replaced by a `1'. All other cells were then set to `0'. The number of common links over various consecutive year periods, was then determined by analyzing these matrices year by year. In particular, common links correspond to 1's in equivalent positions on consecutive matrices. This random voting process was performed 200 times in order to mimic an ensemble average. Then the average and standard deviation of the number of common links for each country, in a particular consecutive year period, was calculated.

The occasional non-attendance of many of the frequently-participating countries in the years between 1992 and 2003, means that while it is often possible to get a reasonable number of results for these countries over two or three consecutive years, for larger periods the results become sparse. Since we are primarily interested in comparing the number of common links over all possible periods (i.e. over multiple timescales) we will focus here on those countries which were present for eleven or twelve consecutive years. This includes the UK, France, Sweden, Spain, Malta and Croatia.
	We find that these countries can be broadly categorized into two groups: those that have a number of common links that is nearly always greater than the random value and those where it is nearly always less than or comparable to the random value. The approximate categorization is as shown in Table 1(d).

This common link analysis reveals a number of new and surprising results. Figure 6 shows examples of the observed and random results for the UK and France, over two and seven consecutive year periods. We find that there are several countries that have a number of common links -- over varying periods of time -- that are consistently higher than the values predicted by the random graph. These include the UK, Sweden, Malta, Ireland, Germany, Cyprus and Estonia (when the existence of common links is not prevented by their non-attendance). 
The existence of countries with observed numbers of common links consistently larger than the `random contest' result, again suggests a degree of cliquishness. These countries seems to possess strong bonds with other competitors, and these bonds provide regular channels for the exchange of points. 
There are also a number of countries, including France, with a consistently low number of common links. These results are particularly noteworthy when considered in conjunction with the fact that there are other countries which possess numbers of common links commensurate with the random regime. 

Most remarkably, it is the UK which seems to possess the highest number of these strong bonds. This finding is at odds with the popular notion (in particular within the UK itself) that the UK is somehow adrift from Europe. Indeed, using the metaphor of the Song Contest itself, this network analysis seems to suggest that {\em the UK is the country which is the most in tune with the rest of Europe} \cite{ref16}. The countries at the opposite end of the spectrum -- France and, to a lesser degree,  Spain -- possess observed numbers of common links that are consistently lower than the `random contest' value, although they do fall within the one standard deviation error bars. This is shown explicitly in Fig. 6 for France. Such a voting pattern could arise if a country has fixed views on the type of song they wish to vote for, but the competitors continually change the nature of their entries. This country will then be forced to vary to whom they assign points, resulting in seemingly random voting behaviour. A surprising suggestion of this analysis is therefore that France is {\em out of tune} with the rest of Europe, since the links which it forms are largely incompatible with the links which it receives back.
	
	More generally, we note that this method of common link analysis -- which we believe  to be new -- should provide a useful tool for determining trends in a general class of evolving networks that do not necessarily have many similarities between consecutive time steps. It should, for example, be directly applicable to determining persistant relationships in the World Trade web \cite{ref8} where the available trading partners are continually changing. We leave this interesting line of investigation for a future paper.

\begin{figure}[ht] 
\includegraphics[width=.54\textwidth]{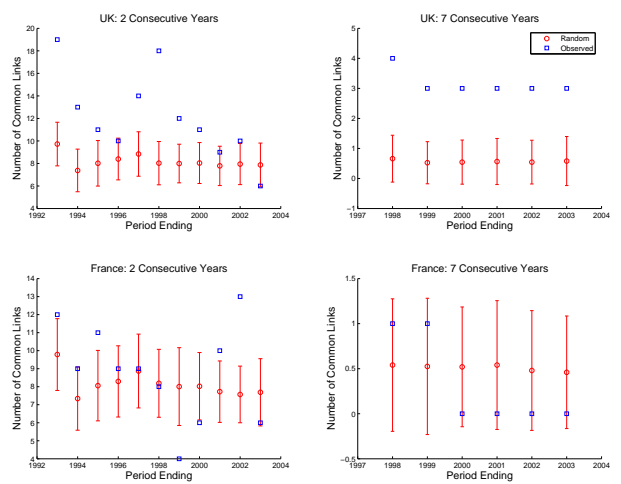}
\caption{The number of common links which are actually observed,  and those expected for a `random contest'. Countries shown are UK (top) and France (bottom) for two (left) and seven (right) consecutive-year periods. The blue points represent the observed values and the red points represent the `random contest' values (shown with error bars of one standard deviation).
}
    \label{fig:figure6} 
\end{figure} 

\vskip0.2in

\noindent{\em{3.7  Reciprocal Links}}

The number of reciprocal links, in which two countries both assign points to each other in a particular year, can also be used as a measure of the cliquishness of countries. In addition, a reciprocal link between two countries could be reasonably interpreted as meaning that they each recognize something positive in the other's song in that particular year. If a particular country possesses many such links, it suggests that their song has a range of similarities with other entries and thus that this country has some deeper understanding of what the rest of Europe appreciates -- it is  again {\em in tune} with Europe. 
	
	We have determined the observed number of reciprocal links for each country through an analysis of the individual country-centred networks (i.e. Fig. 2). The corresponding `random contest' data were found as follows. Let there be $N+1$ countries in a particular year. The probability that a reciprocal link exists between two countries is then given by $(10/N)^2$. 
Hence the expected number of reciprocal links which a country will have in a given year is
$R=(100/N)$. Figures 7 and 8 show the numbers of reciprocal links for the observed data and for a Ôrandom contestÕ, for six regular attendees as a function of time. The graphs tend to be noisier than the common-links graphs of Fig. 6, since they are now responding directly to annual fluctuations. However they still manage to show that each country has years in which it is above or below the expected random number of reciprocal links. It is also possible to identify countries that spend most of their time above the random line, often with values significantly greater than the random values, and those that spend much of their time below the random line. As can be seen, the UK, Sweden and Malta spend more years above the line than below, and in many cases the observed numbers of common links for these countries are considerably larger than the expected random values. In contrast, France and Spain have several years in which they are significantly below the random line.

\begin{figure}[ht] 
\includegraphics[width=.48\textwidth]{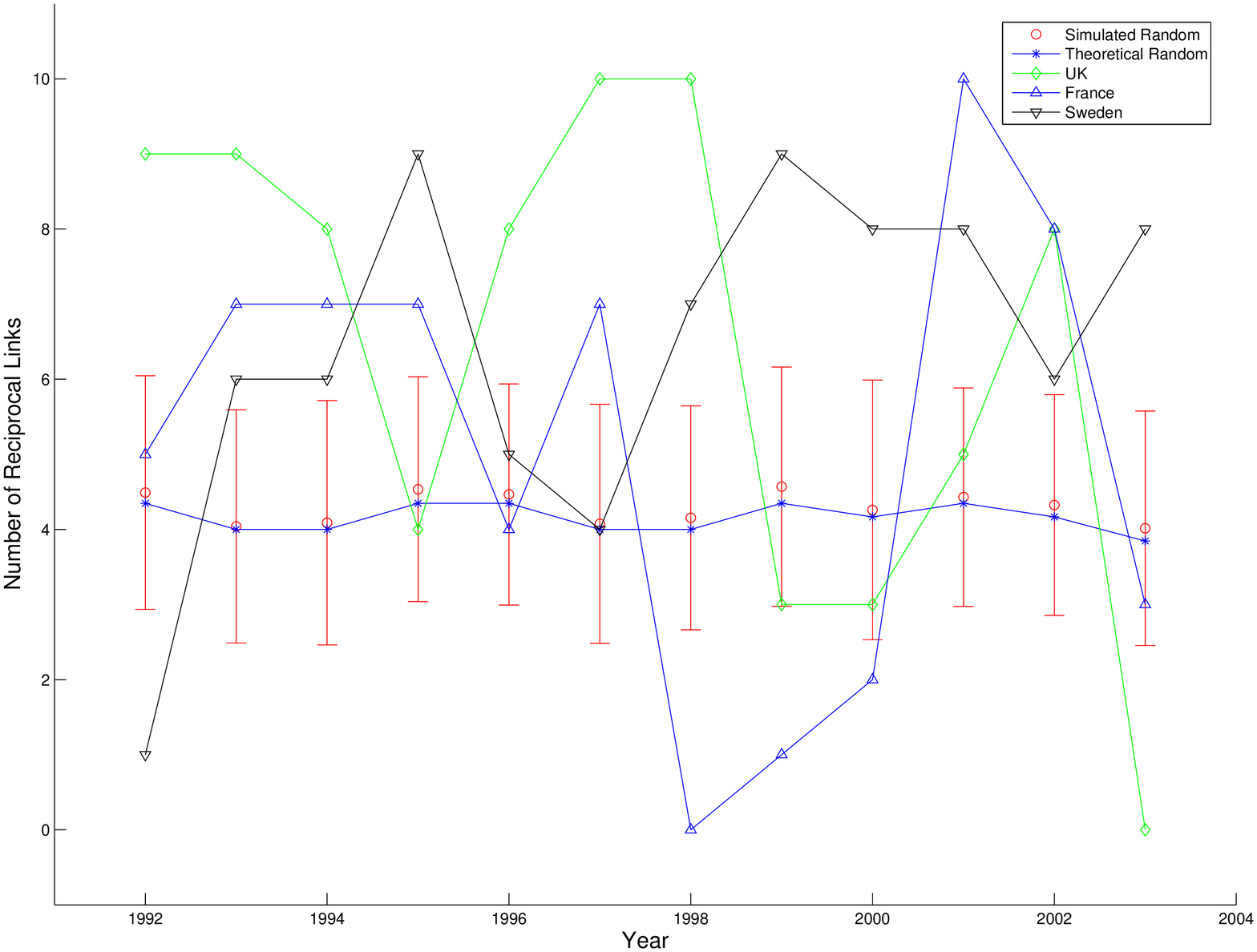}
\caption{Comparison of the observed number of reciprocal links, and the expected number of reciprocal links for a `random contest'. Results are shown for UK, France and Sweden.
}
    \label{fig:figure7} 
\end{figure} 

\begin{figure}[ht] 
\includegraphics[width=.48\textwidth]{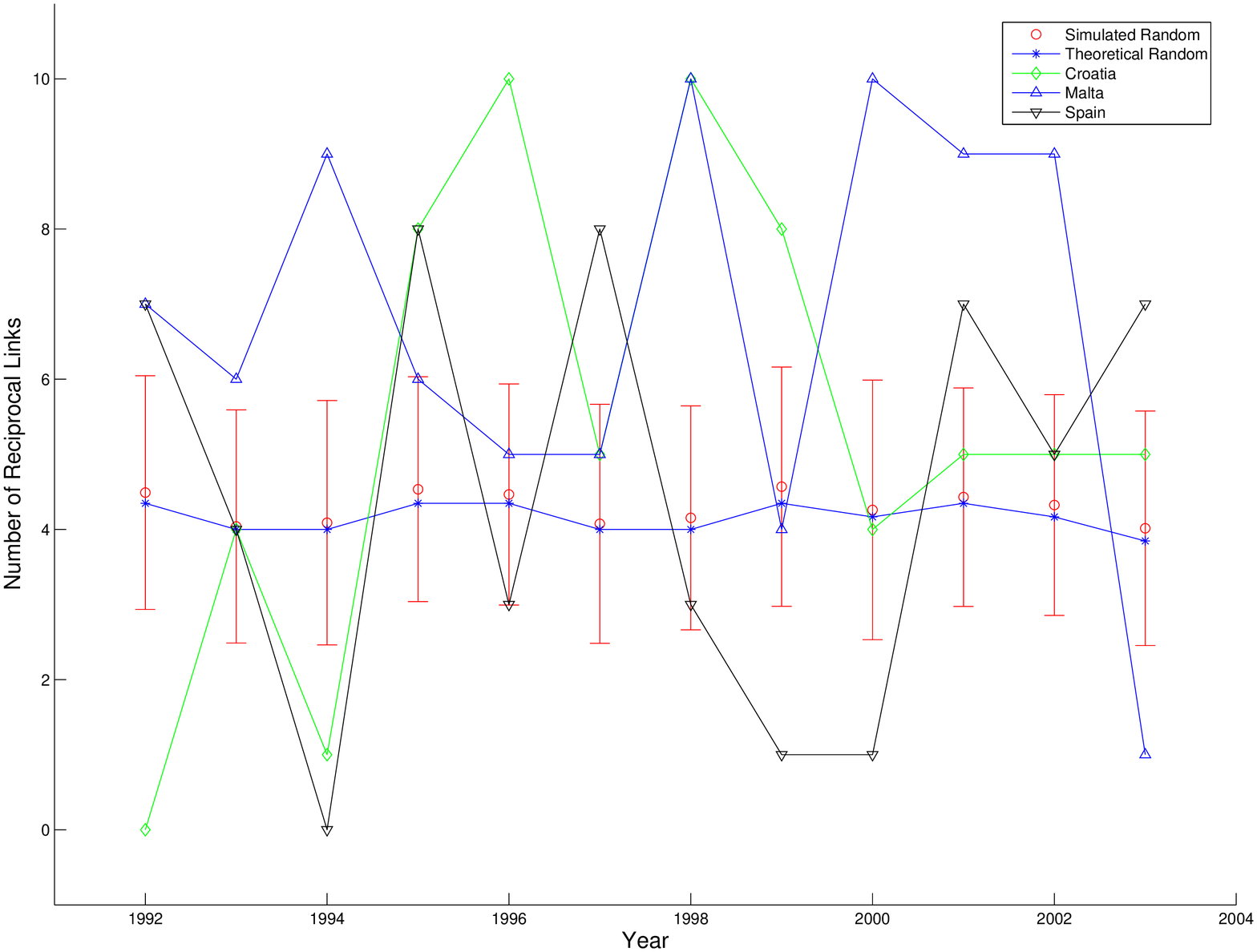}
\caption{Same as Fig. 7, except that the comparison of reciprocal links is now shown for Croatia, Malta and Spain.
}
    \label{fig:figure8} 
\end{figure} 

	These observations are consistent with our earlier finding that countries such as the UK, Sweden and Malta are more in tune with the rest of Europe. As noted earlier, it is particularly surprising that the UK falls within this group since it is renowned for its apparent apathy towards European unity. It is also worth noting that many of the countries identified as having high numbers of reciprocal links also possess large numbers of persistent common links. Countries that fall into both of these categories include the UK, Sweden, Ireland, Malta, Cyprus and Estonia. This may suggest that many of these persistent common links are due to countries with which particular givers hold reciprocal relationships, where these relationships may involve reciprocal behavior which is non-local in time (i.e. stretches over many years). This also supports the idea of voting cliques, with preferential pointsÕ allocation between groups of countries both within a given year, and extending between years.

In Figure 9 we investigate such dynamical effects related to these reciprocal links. In particular, we look into the possibility that there are reciprocal actions over time, possibly even culminating in tit-for-tat type voting behaviour.  We do this by considering two new variables for each country. Consider country A: the variable GR (i.e. Give$_{t-1}$Receive$_t$) is the number of countries for which A voted in contest $t-1$  {\em and} 
which then voted for A in contest $t$.
This attempts to capture the effect whereby countries may reward A in a particular year for votes which A gave to them in the previous year.
Likewise, we introduce a variable RG (i.e. Receive$_{t-1}$Give$_t$) which is the number of countries which voted for A in contest $t-1$ {\em and} for which A voted in contest $t$.

\begin{figure}[ht] 
\includegraphics[width=.54\textwidth]{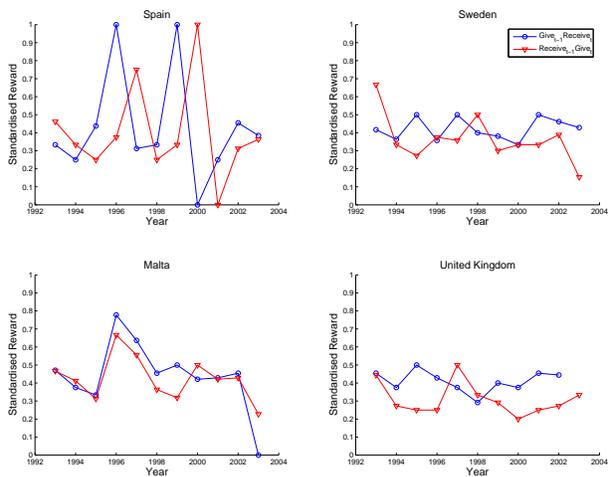}
\caption{Reciprocity of giving and receiving over time, as measured using the variables GR (i.e. Give$_{t-1}$Receive$_t$) and RG (i.e. Receive$_{t-1}$Give$_t$). See text for explanation). We have normalized GR by the (unweighted) votes received in year $t$, and RG by the votes received in year $t-1$. This normalization prevents fluctuations in the number of votes received from dominating the dynamics of the resulting time series. (Note that, as a result of the rules for awarding points, the unweighted votes given out will remain strictly constant over time).
}
    \label{fig:figure9} 
\end{figure} 

The plots of GR and RG for each of the four countries in Fig. 9, appear highly-correlated and may also have a time-lag. The variable GR (i.e. Give$_{t-1}$Receive$_t$) seems
to lead in several cases. 
Malta seems to exhibit very good judgement in terms of giving and receiving points -- in particular, it tends to
give/receive with one group of countries at exactly the same timestep as
it receives/gives with another. Spain seems to have a definite one-year lag while the U.K. reconciles its give/receive and receive/give groups over a two-year timescale.

\vskip0.4in

\noindent{\bf{4. Conclusions}}

We have presented a comprehensive study of a real-world complex network, the `Eurovision points-exchange network' and have uncovered a number of new and surprising trends within the data. As a side-product of this investigation, we have also developed a set of simple yet effective analysis tools which are applicable to finite networks which evolve over time. Indeed, the Eurovision network is analogous to a range of other complex dynamical networks involving the exchange of goods or opinions. Consequently, both the particular study and the tools developed in this paper should be of wider interest to the complex systems and complex networks community.

To tackle the problem of identifying trends within the complex Eurovision exchange system, we investigated both global and local network properties. We began by calculating the clustering coefficients of the yearly networks and the overall degree distribution. These results were found to differ significantly from the results expected for an equivalent random network or `random contest' and thus provided the first suggestion that voting patterns do indeed exist. We then performed a cluster analysis based on the similarities between the number of points that countries allocated to other entrants. The resulting dendrogram highlighted explicitly that voting clusters exist, in particular demonstrating that Cyprus and Greece assigned very similar numbers of points to each of the other countries.
The clusters identified in this way were used as the basis for an analysis of individual nodes. The observed probabilities that connections existed between countries in a cluster, were calculated and found to be significantly greater than the equivalent values for the random contest.  By contrast, the observed probabilities for countries in groups not identified as clusters, were found to be comparable to the random contest results. This again supports the theory that voting cliques exist, although the evidence suggests that they are not based simply on geographical closeness.
	
	Further evidence for clustering was provided by an analysis of the numbers of reciprocal and common links. This analysis helped identify those countries that seem to be most `in tune' with the rest of Europe. Most surprisingly, this study has led to the suggestion that the UK is closely united with the rest of Europe, while France, for example, is not.
	
Although this study concerns, and is strictly limited to, the data emerging from a music competition, we will speculate for the moment on its possible wider context. It is clear that we have uncovered non-trivial and non-random behaviour in the voting dynamics. Implicit in our analysis is the assumption that all songs are, on the face of it, equally `nice' -- in other words, they are of equal musical quality and hence any differences in preferences expressed by a given country A are completely related to the question of `taste'. Underlying such `national taste' is the idea that a country may collectively have some reasonably well-defined preferences -- just like an individual socio-economic agent within the research literature. Indeed, we believe that the spatio-temporal complexity that we have observed in the interactions of our $N$ countries, is not unlike that expected within a group of $N$ interacting human beings -- and in particular, a group of $N$ heterogeneous agents. For this reason, it will be interesting to continue monitoring the Eurovision Song Contest well into the future, in order to monitor possible changes as other `agents' (i.e. countries) enter the arena (i.e. Eurovision Song Contest). It is clear from the rules of the contest, that the complexity which we observe throughout this study arises from the process of group `self-assessment' and that this process, while driven by well-defined local rules, has no central controller. One might therefore speculate that the complexity that we have observed, is representative of many real-world situations in which groups of `agents' establish their own-interrelationships and hence ultimately decide their own fate. Possible examples include groups of individuals, societies, political groups or even governments.

Finally, we acknowledge that there are an infinite number of possible network measures that we could have applied to the present data-set. We also acknowledge that these other network measures might conceivably have led to somewhat different conclusions, although just how different would remain to be seen. However, we would like to emphasize that the present study -- which by definition has had to be finite and limited -- was carried out in an unbiased way in terms of the choice of network tools employed. We chose our set of tools a priori, without having analyzed the details of the data, and this led us to the surprising findings and conclusions which we have reported here. It remains to be seen whether voting behaviour in future Eurovision Song Contests will remain consistent with the findings of this paper. That is part of the interest, to see how things may evolve over time -- and exactly how that happens will depend on how individual countries make their minds up.

\vskip0.2in

\noindent{\bf{Acknowledgements}}

N.J. is grateful to G. Harrison for clarifying information about the changes in rules for voting in past years, and also to F. Reed-Tsochas and other members of CABDyN (Complex Agent-Based Dynamical Networks) at Oxford University for discussions. O.S. holds a research studentship from the Government of Pakistan.


\begin{thebibliography}{99}
\bibitem{ref15} PearsonÕs correlation coefficient is a standard statistical quantity, providing a measure of the degree to which two variables are related. In the present context, it provides a measure of the closeness of two countries by considering the average number of points that both countries donate to each recipient. See, for example, mathworld.wolfram.com/CorrelationCoefficient.html
\end{thebibliography}
\end{document}